\documentclass[a4page,10pt]{article}
\usepackage{amsmath,amssymb,leftidx,cite,epsfig,subfig,graphicx}
\begin{document}

\title{\begin{flushright}
\small IISc/CHEP/01/12
\end{flushright}
\vspace{0.25cm} Signatures of New Physics from HBT Correlations in UHECRs}
\author{Rahul Srivastava\footnote{rahul@cts.iisc.ernet.in}  
\\ \\
\begin{small}Centre for High Energy Physics, Indian
Institute of Science,
Bangalore, 560012, India.
\end{small}}
\date{\empty}

\maketitle

\begin{abstract}
Quantum fields written on noncommutative spacetime (Groenewold - Moyal plane) obey twisted commutation relations. In this paper we show that these twisted commutation relations result
in Hanbury-Brown Twiss (HBT) correlations that are distinct from that for ordinary bosonic or fermionic fields, and hence can provide us useful information about underlying 
noncommutative nature of spacetime. The deviation from usual bosonic/fermionic statistics becomes pronounced at high energies, suggesting that a natural place is to look at
Ultra High Energy Cosmic Rays (UHECRs). Since the HBT correlations are sensitive only to the statistics of the particles, observations done 
with UHECRs are capable of providing unambiguous signatures of noncommutativity, without any detailed knowledge of the mechanism and source of origin of UHECRs. 

\end{abstract}

%%%%%%%%%%%%%%%%%%%%%%%%%%%%%%%%%%%%%%%%%%%%%%%%%%%%%%%%%%%%%%%%%%%%%%%%%%%%%%%%%%%%%%%%%%%%%%%%%%%%%%%%%%%%%%%%%%%%%%%%%%%%%%%%%%%%%%%%%%%%%%%%%%%%%%%%%%%%%%%%%%%%%%%%%%%%%%%%%%%%%%%%%%

\vspace{0.75cm} 

Cosmic Rays with energies around $10^{18}$ eV and higher are called as Ultra High Energy Cosmic Rays (UHECRs) \cite{book,nag,bhat}. They are the highest energy particles known to us 
and often have energies $10^7$ times more than that produced by LHC. Inspite of recent advancements (both theoretical and experimental), UHECRs pose a considerable theoretical 
challenge: the source and mechanism of origin of such high energy particles \cite{pierre4,pierre5} and their composition are areas of active research \cite{pierre2,pierre3,wilk}. \\

Due to their extremely high energies, UHECRs are not only an excellent arena for testing the validity of known laws of physics \cite{coleman,grillo} but are also some 
of the best places to look for signatures, if any, of new physics e.g. theories with Lorentz violation and/or deformed dispersion relations \cite{wolfgang,sigl,luca,stecker,stecker1}.
In this paper we aim to show that UHECRs can be used to look for signatures of a particular model of nonlocalities coming from the underlying noncommutative nature of spacetime at
short distances. \\   

Simple intuitive arguments involving standard quantum mechanics uncertainty relations suggest that at length scales close to Plank length, strong gravity effects will limit the 
spatial as well  as temporal resolution beyond some fundamental length scale (L $\approx$ Planck Length), leading to space - space as well as space - time uncertainties \cite{dop}. 
One cannot probe spacetime with a resolution below this scale i.e. spacetime becomes “fuzzy” below this scale, resulting into noncommutative spacetime. Hence it becomes important 
and interesting to study in detail the structure of such a noncommutative spacetime and the properties of quantum fields written on it, because it not only helps us improve our 
understanding of the Planck scale physics but also helps in bridging standard particle physics with physics at Planck scale . \\

This noncommutative scale need not always be as small as Planck scale. For instance if there are large extra dimensions, then the ``effective Planck scale'' can be at much lower 
energies (usually taken between TeV scale and GUT scale, depending on specific models), resulting in appearance of noncommutativity at scales much larger than Planck scale. 
Such large scale noncommutativity is of particular interest as its effect can be detected in present day or near future experiments. In this paper we are interested in looking 
for signatures of such large scale noncommutativity in UHECRs.\\   

Of the various approaches to model the noncommutative structure of spacetime, the simplest is one  where the coordinates satisfy commutation relations of the form

\begin{eqnarray} 
[\hat{x}_{\mu},\hat{x}_{\nu}] = i \theta_{\mu\nu} \; ; \qquad  \mu, \nu \, = \, 0,1,2,3  \qquad \text{and} \qquad   \theta \: \; \text{is a real, constant, antisymmetric matrix.}
\label{gm}
\end{eqnarray} 

The elements of the $\theta$ matrix have the dimension of $\text{(length)}^2$ and set the scale for the area of the smallest possible localization in the 
$\mu - \nu$ plane, giving a measure for the strength of noncommutativity \cite{dop1}. The algebra generated by $\hat{x}_{\mu}$ is usually referred to as Groenewold - Moyal (G.M) plane 
\cite{balreview}, and in this paper we will restrict our attention to this noncommutative spacetime. Equivalently this noncommutative nature of spacetime can be taken into account by 
defining a new type of multiplication rule ($\ast$ product) to multiply functions evaluated at same point i.e. 

 \begin{eqnarray} 
 f(x) \ast g(x) = f(x) e^{\frac{i}{2}\overleftarrow{\partial}_{\mu} \theta^{\mu\nu} \overrightarrow{\partial}_{\nu} } g(x).
\end{eqnarray} 

One particularly important feature of G.M. plane, which makes it quite suitable for writing quantum field theories on it, is the restoration of Poincar\'e-Hopf symmetry
as Hopf algebraic symmetry, by defining a new coproduct (twisted coproduct) for action of Poincar\'e group elements on state vectors \cite{wess} \cite{cha} \cite{cha1}.\\

Twisting of the coproduct has immediate implications for the symmetries of multi-particle wave functions describing identical particles \cite{sachin}. For example, on G.M plane
 the correct physical two-particle wave functions are twisted (anti-)symmetrized ones and are given by

\begin{eqnarray} 
  \phi \otimes_{S_{\theta}} \psi & \equiv &  \left( \frac{1 \, + \, \tau_{\theta}}{2} \right) ( \phi \otimes \psi ) \nonumber \\
\phi \otimes_{A_{\theta}} \psi & \equiv &  \left( \frac{1 \, - \, \tau_{\theta}}{2} \right) ( \phi \otimes \psi )
\label{tsym}
 \end{eqnarray}

where $\phi$ and $\psi$ are single particle wavefunctions (of two identical particles) and $\tau_{\theta}$ is the twisted statistics (flip) operator associated with exchange given by 

\begin{eqnarray}
 \tau_{\theta} & = & \mathcal{F}^{-1} \tau_0 \mathcal{F}, \qquad \qquad \mathcal{F} \; = \; e^{\frac{i}{2} \theta^{\mu\nu} \partial_{\mu} \otimes \partial_{\nu} } .
 \label{tflip}
\end{eqnarray}

Here $\tau_0$ is the commutative flip operator : $\tau_0 \, ( \phi \otimes \psi ) \, = \,  \psi \otimes \phi$. \\

The above analysis can also be extended to field theories on G.M plane resulting in changed commutation relations between creation/annihilation operators \cite{sachin}, which now become

\begin{eqnarray} \left.
\begin{array}{l l l}
a_{\bold{p}}a_{\bold{q}} &=& \eta e^{ip\wedge q}a_{\bold{q}}a_{\bold{p}}, \quad a^{\dagger}_{\bold{p}}
a^{\dagger}_{\bold{q}} = \eta e^{ip\wedge q}a^{\dagger}_{\bold{q}}a^{\dagger}_{\bold{p}} \\
a_{\bold{p}}a^{\dagger}_{\bold{q}} &=& \eta e^{-ip\wedge q}a^{\dagger}_{\bold{q}}a_{\bold{p}} + 
(2\pi)^{3}2p_{0} \delta^{3}(\bold{p}-\bold{q}) 
\end{array}\right\}
\begin{array}{l l l}
\qquad \text{where} \quad p^{\mu} = (p^{0}, \bold{p}), \quad p\wedge q = p_\mu \theta^{\mu\nu}q_\nu, \\
\qquad \text{and} \quad \eta = \pm 1  \quad \text{for bosons/fermions.}
\end{array}
\label{tcom}
\end{eqnarray}

In literature there exist another approach to quantization of noncommutative fields \cite{chaic}. In this approach, the quantization is done according to the usual rules and the 
quantum fields follow usual bosonic/fermionic statistics. However, such a quantization scheme does not preserve the classical twisted Poincar\'e invariance and suffers
from UV/IR mixing \cite{pinzul} . In this paper we only discuss the twisted quantization as it preserves twisted Poincar\'e invariance in noncommutative quantum field theories. \\
 
Because of (\ref{tcom}) the quantum fields written on G.M plane, unlike ordinary quantum fields, follow a unusual statistics which we call as twisted statistics. Twisted statistics 
are a unique feature of fields on G.M plane and can be used to search for signals of noncommutativity: because of the twisted commutation relations, interesting new effects like 
Pauli forbidden transitions \cite{bal,pramod} can arise. The effect of twisted statistics also manifests itself in certain thermodynamic quantities \cite{basu}, \cite{basu1}.  
The two-point distribution functions remain unchanged

\begin{eqnarray} 
 \left\langle a^{\dagger}_{\bold{p_{1}}}a_{\bold{p_{2}}}\right\rangle & = & 2\left(p_{1}\right)_{0} \, N^{(T)}_{\bold{p_{1}}} \, \delta^{3}(\bold{p_{1}}-\bold{p_{2}}) 
\label{2cor}
\end{eqnarray}

where $ N^{(T)}_{\bold{p}} =  \frac{1}{e^{\beta E_{\bold{p}}}-\eta}$ is the thermal distribution, but for example the quantity 
$ \left\langle a^{\dagger}_{\bold{p_{1}}} a^{\dagger}_{\bold{p_{2}}} a_{\bold{p_{3}}}a_{\bold{p_{4}}}\right\rangle$ gets changed \cite{basu}.
 
\begin{eqnarray} 
\left\langle a^{\dagger}_{\bold{p_{1}}} a^{\dagger}_{\bold{p_{2}}} a_{\bold{p_{3}}}a_{\bold{p_{4}}}\right\rangle 
= 2\left(p_{1}\right)_{0} 2\left(p_{2}\right)_{0} N^{(T)}_{\bold{p_{1}}} N^{(T)}_{\bold{p_{2}}} \left[ \delta^{3}(\bold{p_{1}}-\bold{p_{4}})\delta^{3}(\bold{p_{2}}-\bold{p_{3}}) 
+ \eta e^{i p_{1} \wedge p_{2}} \delta^{3}(\bold{p_{1}}-\bold{p_{3}})\delta^{3}(\bold{p_{2}}-\bold{p_{4}})\right]
\label{4cor}
\end{eqnarray} 

The above differs from the commutative expression by the appearance of the factor $ e^{i p_{1} \wedge p_{2}} $ in the second term.\\

One can easily check, following a analysis similar to that done in \cite{basu}, that (\ref{2cor}) and (\ref{4cor}) are true not only for thermal distribution  $ N^{(T)}_{\bold{p}}$ 
but for any arbitrary wavepacket $ f(\bold{p})$ i.e. 

\begin{eqnarray} 
  \left\langle a^{\dagger}_{\bold{p_{1}}}a_{\bold{p_{2}}}\right\rangle & = &  2\left(p_{1}\right)_{0} \, f(\bold{p_{1}}) \, \delta^{3}(\bold{p_{1}}-\bold{p_{2}}) 
\label{g2cor}
\end{eqnarray} 
 
and

\begin{eqnarray} 
 \left\langle a^{\dagger}_{\bold{p_{1}}} a^{\dagger}_{\bold{p_{2}}} a_{\bold{p_{3}}}a_{\bold{p_{4}}}\right\rangle 
\, = \, 2\left(p_{1}\right)_{0} 2\left(p_{2}\right)_{0} f(\bold{p_{1}}) f(\bold{p_{2}}) \left[ \delta^{3}(\bold{p_{1}}-\bold{p_{4}})\delta^{3}(\bold{p_{2}}-\bold{p_{3}}) 
+ \eta e^{i p_{1} \wedge p_{2}} \delta^{3}(\bold{p_{1}}-\bold{p_{3}})\delta^{3}(\bold{p_{2}}-\bold{p_{4}})\right] .
\label{g4cor}
\end{eqnarray}

Here we discuss the consequences of (\ref{g4cor}) to the HBT correlation functions.\\

Hanbury-Brown Twiss (HBT) effect \cite{han} is the interference effect between intensities measured by two detectors when a beam of identical particles is projected on them, 
with the intensities recorded by the two detectors operating simultaneously. This intensity increases for bosons (and decreases for fermions) when compared with the intensities recorded 
by the same two detectors, if only one is operated at a time.  The correlation function for HBT effect is defined as

\begin{eqnarray} 
C = \frac{\langle I_{1}\cdot I_{2}\rangle}{\langle I^{'}_{1}\rangle \langle I^{'}_{2}\rangle} \; ; \qquad \qquad \text{where}
\end{eqnarray} 

$ I_{1} $, $ I_{2} $  $=$ intensities recorded by the two detectors respectively, when both are operated simultaneously.\\
$ I^{'}_{1} $ $=$  intensity recorded by first detector when the second detector is not operating. \\
$ I^{'}_{2} $ $=$  intensity recorded by second detector when the first detector is not operating. \\

The HBT correlation function $C$ obeys

\begin{eqnarray} 
 & & C = 1 \qquad \text{for distinguishable particles} \nonumber\\
& & C > 1 \qquad \text{for bosons (bunching effect)} \nonumber\\
& & C < 1 \qquad \text{for fermions (anti-bunching effect)}
\end{eqnarray} 

In the commutative case, for a beam of identical (massless, scalar \footnote{ Scalar bosons are taken for sake of simplicity but the analysis presented here can be easily generalized 
to higher spin bosons.}) bosons, the HBT correlation function can be written as \cite{glauber,glauber1}

\begin{eqnarray} 
C^{(B)}_{0} =  \frac{\left\langle \phi^{(-)}_{0}(\bold{y_{1}})\phi^{(-)}_{0}(\bold{y_{2}})\phi^{(+)}_{0}(\bold{y_{2}})\phi^{(+)}_{0}(\bold{y_{1}}) \right\rangle}
{\left\langle \phi^{(-)}_{0}(\bold{y_{1}})\phi^{(+)}_{0}(\bold{y_{1}}) \right\rangle \; 
\left\langle \phi^{(-)}_{0}(\bold{y_{2}})\phi^{(+)}_{0}(\bold{y_{2}}) \right\rangle}
\label{ccorb}
\end{eqnarray} 

where $\bold{y_{1}}$ and $\bold{y_{2}}$ are the position of the two detectors, $\phi^{(+)}_{0}(\bold{y}) = \int \frac{d^{3}\bold{p}}{(2\pi)^{3}} \, \frac{1}{2E_{\bold{p}}} \,
c_{p} \, e^{i\bold{p}\cdot\bold{y}} $ is the positive frequency part and
 $\phi^{(-)}_{0}(\bold{y}) = \int \frac{d^{3}\bold{p}}{(2\pi)^{3}} \, \frac{1}{2E_{\bold{p}}} \, c^{\dagger}_{p} \, e^{-i\bold{p}\cdot\bold{y}} $ is the negative frequency part of 
the bosonic quantum field $\phi_{0}$ \footnote{ We denote the usual bosonic (fermionic) fields by $\phi_{0}$ ($\psi_{0}$) and their 
twisted counterparts are denoted by $\phi_{\theta}$ ($\psi_{\theta}$). Also, the usual creation/annihilation operators are denoted by $c^{\dagger}_p$ 
and $c_p$ whereas the twisted ones are denoted by $a^{\dagger}_p$ and $a_p$ respectively. }. Notice that we have deliberately taken massless fields because we are interested
in looking for HBT correlation functions of ultra-relativistic particles. \\

To account for the uncertainties in the energy measurements, which at present are quite significant for UHECRs, the incoming beam should be taken as a wavepacket, 
instead of plane waves. The choice of an appropriate wavepacket is potentially the only place where the information about the details of production mechanism or source of origin of 
UHECRs can reside. Central Limit Theorem tells us that the mean of a sufficiently large number of independent random variables, each with finite mean and 
variance, will be approximately distributed like a Gaussian, and hence it is a good first approximation to take the wavepacket as a Gaussian wavepacket 
$ f(\bold{p}) = N e^{- \alpha(\bold{p}- \bold{p_{0}})^{2}}$ centered around some mean momentum $\bold{p_{0}}$. \\

Taking  the wavepacket to be this Gaussian, restricting ourself to only coincidence measurements and using the standard integrals \cite{grad}, the correlation function turns out to be 

\begin{eqnarray}
C^{(B)}_{0} = 1 + e^{- \frac{ \bold{y}^{2}}{2\alpha}}
\label{corgaus}
\end{eqnarray}

As clear from (\ref{corgaus}), $C^{(B)}_{0}$ depends only on  y $ = |\bold{y_{1}} - \bold{y_{2}}|$ the separation between detectors and on $\alpha $. There is no dependence on the mean
 momentum $\bold{p_{0}}$ \cite{baym,baymbook}. \\

Similarly, for a beam of identical (chiral) fermions, the HBT correlation function can be written as 

\begin{eqnarray} 
C^{(F)}_{0} = \frac{\left\langle \overline{\psi}^{(-)}_{0}(\bold{y_{1}})\gamma_{0}\overline{\psi}^{(-)}_{0}(\bold{y_{2}})\gamma_{0}\frac{1}{2}(1\pm \gamma_{5})\psi^{(+)}_{0}(\bold{y_{2}})
\frac{1}{2}(1\pm \gamma_{5})\psi^{(+)}_{0}(\bold{y_{1}}) \right\rangle}{\left\langle\overline{\psi}^{(-)}_{0}(\bold{y_{1}})
\gamma_{0}\frac{1}{2}(1\pm \gamma_{5})\psi^{(+)}_{0}(\bold{y_{1}}) \right\rangle \left\langle \overline{\psi}^{(-)}_{0}(\bold{y_{2}})\gamma_{0}
\frac{1}{2}(1\pm \gamma_{5})\psi^{(+)}_{0}(\bold{y_{2}}) \right\rangle}
\label{ccorf}
\end{eqnarray} 

where, as before, $\bold{y_{1}}$ and $\bold{y_{2}}$ are the position of the two detectors, $\psi^{(+)}_{0}(\bold{y}) = \int \frac{d^{3}\bold{p}}{(2\pi)^{3}} \, \frac{1}{2E_{\bold{p}}}
 \,\sum_{s} \, u_{s, \bold{p}} \,c_{s,p} \, e^{i\bold{p}\cdot\bold{y}} $ is the positive frequency part and $\overline{\psi}^{(-)}_{0}(\bold{y}) = \int \frac{d^{3}\bold{p}}{(2\pi)^{3}}
 \, \frac{1}{2E_{\bold{p}}} \, \sum_{s} \, \overline{u}_{s, \bold{p}}\, c^{\dagger}_{s,p} \, e^{-i\bold{p}\cdot\bold{y}} $  is the negative frequency part of the fermionic quantum 
field $\psi_{0}$ \\

Since the whole analysis is done keeping ultra-relativistic particles in back of our mind, here we look for correlation between only chiral fermions (left-left or right-right): 
at such high energies, the particles are effectively massless and hence chiral fermions are the more appropriate ones to deal with \footnote{ $C_0^{(F)} = 1$ between particles 
with opposite helicities (i.e. between left-right or right-left), as at ultra-relativistic energies, they behave like distinguishable particles}.\\

Taking the incoming beam as a Gaussian wavepacket $ f(\bold{p}) = N e^{- \alpha(\bold{p}- \bold{p_{0}})^{2}}$ centered around some mean momentum $\bold{p_{0}}$ and restricting ourself
 to only coincidence measurements, the correlation function turns out to be 

\begin{eqnarray} 
 C^{(F)}_{0} & = & 1  - \frac{e^{-\frac{\bold{y}^{2}}{2\alpha}}}{2}  - \frac{2}{9 \pi \alpha} e^{-2\alpha \bold{p_{0}}^{2}}\left(\bold{y}^{2} + 4\alpha^{2}\bold{p_{0}^{2}}\right)    
 \leftidx{_1}{F}{_1} \left[2;\frac{5}{2};\frac{-1}{4\alpha}\left(\bold{y}^{2} - 4\alpha^{2}\bold{p_{0}}^{2} - 4i\alpha\bold{y}\cdot\bold{p_{0}} \right)\right] \nonumber \\
& & \leftidx{_1}{F}{_1} \left[2;\frac{5}{2};\frac{-1}{4\alpha}\left(\bold{y}^{2} - 4\alpha^{2}\bold{p_{0}}^{2} + 4i\alpha\bold{y}\cdot\bold{p_{0}} \right)\right]
\label{comcor}
\end{eqnarray} 

where $\bold{y} = \bold{y_{2}} -\bold{y_{1}} $ is the separation between the two detectors and $ \leftidx{_1}{F}{_1}\left(\alpha;\gamma;z\right)$ is the degenerate hypergeometric function. \\

We observe that in this case, the correlation function depends on the separation between detectors $\bold{y} $, on the mean momentum $ \bold{p_{0}}$ of the wavepacket and the angles
 between $\bold{y} $ and $ \bold{p_{0}}$.\\

In the noncommutative case, there are two important differences. Firstly, the ($\cdot$) product between fields evaluated at the same point has to be replaced by ($\ast$) 
product and secondly the expectation value is changed in accordance with (\ref{4cor}), as the quantum fields are now composed of twisted creation/annihilation operators. 
Hence in noncommutative case for twisted (massless, scalar) bosonic particles, we have 

\begin{eqnarray}
C^{(B)}_{\theta} = \frac{\left\langle \phi^{(-)}(\bold{y_{1}})\left \{ \phi^{(-)}(\bold{y_{2}})\ast_{\bold{y_{2}}}\phi^{(+)}(\bold{y_{2}}) \right \}
\ast_{\bold{y_{1}}}\phi^{(+)}(\bold{y_{1}}) \right\rangle}
{\left\langle \phi^{(-)}(\bold{y_{1}})\ast_{\bold{y_{1}}}\phi^{(+)}(\bold{y_{1}}) \right\rangle 
 \left\langle \phi^{(-)}(\bold{y_{2}})\ast_{\bold{y_{2}}}\phi^{(+)}(\bold{y_{2}}) \right\rangle}
\label{ncorb}
\end{eqnarray} 

where $ \ast_{\bold{y}} = e^{\frac{i}{2}\left( \overleftarrow{\partial_{y}}\right)_{\mu}\theta^{\mu\nu}\left( \overrightarrow{\partial_{y}}\right)_{\nu}}$. \\

In rest of the paper we restrict to only space-space noncommutativity \footnote{We have assumed only space-space noncommutativity for calculational simplicity but one can 
do similar analysis with both space-time as well as space-space noncommutativity} i.e. we take $\theta_{0i} = \theta_{i0} = 0$. As in commutative case, we take the incoming beam
 as a Gaussian wavepacket $ f(\bold{p}) = N e^{- \alpha(\bold{p}- \bold{p_{0}})^{2}}$ centered around some mean momentum $\bold{p_{0}}$ and restrict ourselves to only coincidence
 measurements. $C^{(B)}_{\theta}$ then turns out to be 

\begin{eqnarray}
C^{(B)}_{\theta} = 1 + \frac{4\alpha}{4\alpha^{2}+\lambda^{2}} \, \exp{\left[-\frac{4\alpha^{2}\bold{y}^{2} - (\bold{y}\cdot\bold{{\lambda}})^{2}}{2\alpha(4\alpha^{2}+\lambda^{2})}\right]}
\, \exp{\left[-\frac{2\alpha \{\bold{p_{0}}^{2}\bold{\lambda}^{2}-(\bold{p_{0}}\cdot\bold{\lambda})^{2}\}}{4\alpha^{2}+\lambda^{2}}\right]}
\, \exp{\left[-\frac{4\alpha\bold{y}\cdot(\bold{p_{0}}\times\bold{\lambda})}{4\alpha^{2}+\lambda^{2}}\right]}
\label{gcor}
\end{eqnarray}

where we have defined $ \theta_{ij} = \varepsilon_{ijk}\lambda_{k} $. In getting (\ref{gcor}) we have used the standard result that, for $n$-dim column matrices X and  B and a 
$n\times n$ positive definite, symmetric square matrix A , we have 

\begin{eqnarray}
\int d^{n}X_{i} e^{-X^{T}AX + B^{T}X}= \left(\frac{\pi^{n}}{\det{A}}\right)^{\frac{1}{2}}e^{\frac{1}{4}B^{T}A^{-1}B}
\end{eqnarray}

We see that (\ref{gcor}) not only depends on $\bold{y}$ (the separation between detectors) but 
also on the mean momentum $\bold{p_{0}}$, the noncommutative length scale $\bold{\lambda}$, as well as on the angles between $\bold{\lambda}$ and $\bold{y}$ and 
between $\bold{\lambda}$ and $\bold{p_{0}}$. Moreover, as a check we can see that, in the limit $\bold{\lambda}\rightarrow 0$ we get back (\ref{corgaus}).\\

Similarly, for twisted (chiral) fermionic particles the noncommutative HBT correlation function is given by 

\begin{eqnarray}
C^{(F)}_{\theta} = \frac{\left\langle \overline{\psi}^{(-)}(\bold{y_{1}})\gamma_{0} \left \{\overline{\psi}^{(-)}(\bold{y_{2}})\gamma_{0} 
\ast_{\bold{y_{2}}}\frac{1}{2}(1\pm \gamma_{5})\psi^{(+)}(\bold{y_{2}}) \right \}\ast_{\bold{y_{1}}} \frac{1}{2}(1\pm \gamma_{5})\psi^{(+)}(\bold{y_{1}})  \right\rangle}
{\left\langle\overline{\psi}^{(-)}(\bold{y_{1}})\gamma_{0}\ast_{\bold{y_{1}}} \frac{1}{2}(1\pm \gamma_{5})\psi^{(+)}(\bold{y_{1}}) \right\rangle
 \left\langle \overline{\psi}^{(-)}(\bold{y_{2}})\gamma_{0} \ast_{\bold{y_{2}}}\frac{1}{2}(1\pm \gamma_{5})\psi^{(+)}(\bold{y_{2}}) \right\rangle}
\label{ncor}
\end{eqnarray} 

Again restricting ourselves to only space-space noncommutativity, using Gaussian wavepackets and considering only coincidence measurements, $C^{(F)}_{\theta}$  can be written as

\begin{eqnarray}
C^{(F)}_{\theta} & = & 1 - \frac{1}{2} e^{-2\alpha\bold{p_{0}}^{2}} \left[ e^{-\frac{\bold{z_{1}}^{2}}{4\alpha}} e^{i\left( \overleftarrow{\partial_{z_{1}}}\right)_{i}\theta^{ij}
\left( \overrightarrow{\partial_{z_{2}}}\right)_{j}} e^{-\frac{\bold{z_{2}}^{2}}{4\alpha}} \right] - \left( \frac{2 }{9\pi\alpha} \right) e^{-2\alpha\bold{p_{0}}^{2}} 
\left[\left\{ \leftidx{_1}{F}{_1}\left( 2, \frac{5}{2}, -\frac{\bold{z_{1}}^{2}}{4\alpha} \right) (z_{1})_{a}\right\} \right. \nonumber \\ 
& &  \left.e^{i\left( \overleftarrow{\partial_{z_{1}}}\right)_{i}\theta^{ij}\left( \overrightarrow{\partial_{z_{2}}}\right)_{j}}
 \left\{ \leftidx{_1}{F}{_1}\left( 2, \frac{5}{2}, -\frac{\bold{z_{2}}^{2}}{4\alpha} \right) (z_{2})_{a}\right\} \right ]
\label{gennoncor}
\end{eqnarray} 

where $\bold{z_{1}} = \bold{y} - 2i\alpha\bold{p_{0}} $ and $\bold{z_{2}} = \bold{y} + 2i\alpha\bold{p_{0}} $.\\

Expanding this in $ \theta$ and taking terms upto second order we get

\begin{eqnarray}
& & C^{(F)}_{\theta} = 1 -  \frac{1}{2}  e^{-\frac{\bold{y}^{2}}{2\alpha}}  \left[1 + \frac{\bold{p_{0}}\cdot(\bold{y}\times\bold{\lambda})}{\alpha}
 - \frac{\bold{\lambda}^{2}}{4\alpha^{2}} + \frac{\bold{y}^{2}\bold{\lambda}^{2}}{8\alpha^{3}} - \frac{(\bold{y}\cdot\bold{\lambda})^{2}}{8\alpha^{3}} 
 -\frac{\bold{p_{0}}^{2}\bold{\lambda}^{2}}{2\alpha} +  \frac{(\bold{p_{0}}\cdot\bold{\lambda})^{2}}{2\alpha}
 + \frac{[\bold{p_{0}}\cdot(\bold{y}\times\bold{\lambda})]^{2}}{2\alpha^{2}} \right] \nonumber \\
&  & -  \left( \frac{2 }{9\pi\alpha} \right)  e^{-2\alpha\bold{p_{0}}^{2}} \left[  \leftidx{_1}{F}{_1}\left( 2; \frac{5}{2};-\frac{1}{4\alpha}\left( \bold{y}^{2} - 4\alpha^{2}\bold{p_{0}}^{2} 
- 4i\alpha\bold{y}\cdot\bold{p_{0}} \right) \right)  \leftidx{_1}{F}{_1}\left( 2; \frac{5}{2};-\frac{1}{4\alpha}\left( \bold{y}^{2} - 4\alpha^{2}\bold{p_{0}}^{2}
 + 4i\alpha\bold{y}\cdot\bold{p_{0}} \right) \right) \right.  \nonumber \\
& &  \left( \bold{y}^{2} + 4\alpha^{2}\bold{p_{0}}^{2} \right) +  \frac{4}{25\alpha^{2}} \left\{ ( 4 \alpha\bold{p_{0}}\cdot(\bold{y}\times\bold{\lambda}) 
- 2\bold{\lambda}^{2} ) \left( \bold{y}^{2} + 4\alpha^{2}\bold{p_{0}}^{2} \right)   + (\bold{y}\cdot\bold{\lambda})^{2} 
+ 4\alpha^{2}(\bold{p_{0}}\cdot\bold{\lambda})^{2} \right\} \nonumber \\
& &  \leftidx{_1}{F}{_1}\left( 3; \frac{7}{2};-\frac{1}{4\alpha}\left( \bold{y}^{2} - 4\alpha^{2}\bold{p_{0}}^{2} - 4i\alpha\bold{y}\cdot\bold{p_{0}} \right) \right) 
 \leftidx{_1}{F}{_1}\left( 3; \frac{7}{2};-\frac{1}{4\alpha}\left( \bold{y}^{2} - 4\alpha^{2}\bold{p_{0}}^{2} + 4i\alpha\bold{y}\cdot\bold{p_{0}} \right) \right) \nonumber \\
& & + \frac{6}{175 \alpha^{3}} \left\{ \bold{y}^{2}\bold{\lambda}^{2} - 4\alpha^{2}\bold{p_{0}}^{2}\bold{\lambda}^{2}  - 4i\alpha(\bold{p_{0}}\cdot\bold{y})\bold{\lambda}^{2}
 - (\bold{y}\cdot\bold{\lambda})^{2} + 4\alpha^{2}(\bold{p_{0}}\cdot\bold{\lambda})^{2} + 4i\alpha(\bold{p_{0}}\cdot\bold{\lambda})(\bold{y}\cdot\bold{\lambda}) \right\} \nonumber \\
& & ( \bold{y}^{2} + 4\alpha^{2}\bold{p_{0}}^{2} )\leftidx{_1}{F}{_1}\left( 4; \frac{9}{2};-\frac{1}{4\alpha}\left( \bold{y}^{2} 
- 4\alpha^{2}\bold{p_{0}}^{2} - 4i\alpha\bold{y}\cdot\bold{p_{0}} \right) \right)    
\leftidx{_1}{F}{_1}\left( 3; \frac{7}{2};-\frac{1}{4\alpha}\left( \bold{y}^{2} - 4\alpha^{2}\bold{p_{0}}^{2} + 4i\alpha\bold{y}\cdot\bold{p_{0}} \right) \right) \nonumber \\
& &  + \frac{6}{175 \alpha^{3}} \{ \bold{y}^{2}\bold{\lambda}^{2} - 4\alpha^{2}\bold{p_{0}}^{2}\bold{\lambda}^{2}  + 4i\alpha(\bold{p_{0}}\cdot\bold{y})\bold{\lambda}^{2}
 - (\bold{y}\cdot\bold{\lambda})^{2} + 4\alpha^{2}(\bold{p_{0}}\cdot\bold{\lambda})^{2} - 4i\alpha(\bold{p_{0}}\cdot\bold{\lambda})(\bold{y}\cdot\bold{\lambda}) \} \nonumber \\
 & &  ( \bold{y}^{2} + 4\alpha^{2}\bold{p_{0}}^{2} )  \leftidx{_1}{F}{_1}\left( 3; \frac{7}{2};-\frac{1}{4\alpha}\left( \bold{y}^{2} - 4\alpha^{2}\bold{p_{0}}^{2}
 - 4i\alpha\bold{y}\cdot\bold{p_{0}} \right) \right) \nonumber
\end{eqnarray}
\begin{eqnarray}
& &  \leftidx{_1}{F}{_1}\left( 4; \frac{9}{2};-\frac{1}{4\alpha}\left( \bold{y}^{2} - 4\alpha^{2}\bold{p_{0}}^{2} + 4i\alpha\bold{y}\cdot\bold{p_{0}} \right) \right) 
- \frac{288}{1225\alpha^{2}} \{ \bold{p_{0}}\cdot(\bold{y}\times\bold{\lambda}) \}^{2}  ( \bold{y}^{2} + 4\alpha^{2}\bold{p_{0}}^{2} ) \nonumber \\
& & \left. \leftidx{_1}{F}{_1}\left( 4; \frac{9}{2};-\frac{1}{4\alpha}\left( \bold{y}^{2} - 4\alpha^{2}\bold{p_{0}}^{2} - 4i\alpha\bold{y}\cdot\bold{p_{0}} \right) \right)  
\leftidx{_1}{F}{_1}\left( 4; \frac{9}{2};-\frac{1}{4\alpha}\left( \bold{y}^{2} - 4\alpha^{2}\bold{p_{0}}^{2} + 4i\alpha\bold{y}\cdot\bold{p_{0}} \right) \right) \right] + \; O(\lambda^{3})
\label{noncor}
\end{eqnarray}

where again in writing (\ref{noncor}) we have defined $ \theta_{ij} = \varepsilon_{ijk}\lambda_{k} $.\\

As can be clearly seen from (\ref{noncor}), the noncommutative HBT correlation function not only depends on $\bold{y} $ and $ \bold{p_{0}}$ but also on $\bold{\lambda}$ 
and various angles that $\bold{\lambda}$ makes with $\bold{y} $ and $ \bold{p_{0}}$. \\

A comparison of (\ref{corgaus})-(\ref{gcor}) and (\ref{comcor})-(\ref{noncor}) tells us that the HBT correlation function for twisted bosons/fermions are different from 
those for ordinary bosons/fermions and become more pronounced with increasing momenta of incident particles. Since $\bold{\lambda}$ is expected to be a very small quantity,
the deviations will be highly suppressed and one has to look at particles with high enough energy-momentum (w.r.t noncommutative scale), so that the deviations get sufficiently 
enhanced to be detectable. Hence, as stated in beginning of the paper, the best place to look for the signatures of twisted statistics, is to look at particles 
in L.H.C or in Ultra High Energy Cosmic Rays (UHECRs). Moreover, since HBT correlations are essentially quantum correlations between free identical particles and are manifestations 
of the particular statistics followed by the identical particles of a given beam, they are sensitive only to the statistics obeyed by these particles. Therefore despite our 
present lack of detailed knowledge about production mechanism and source of origin of UHECRs, a study of HBT correlations may still be able to provide unambiguous signatures, 
of underlying noncommutative nature of spacetime. \\

The graphs shown below highlight the difference between $C^{(B)}_{0}$ - $C^{(B)}_{\theta}$ and $C^{(F)}_{0}$ - $C^{(F)}_{\theta}$ \footnote{The persistence of correlations
 to large distances is attributed to the large uncertainties in our present determination of energy-momentum of UHECRs. With better and more precise knowledge of the energy-momentum,
 the correlation will decrease significantly resulting in better bounds on noncommutative deviations}. \\

\begin{figure} [h]
\centering
\subfloat []{\includegraphics[angle=0,width=2.1in]{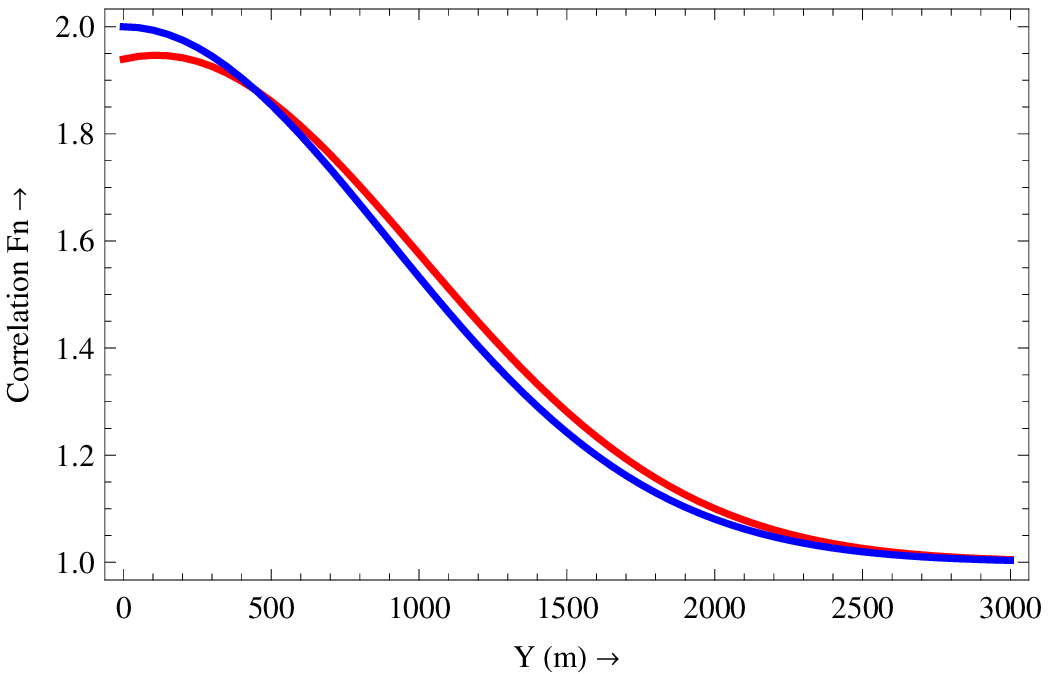} } 
 \hspace{.75cm}\vspace{.15cm}
\subfloat []{\includegraphics[angle=0,width=2.1in]{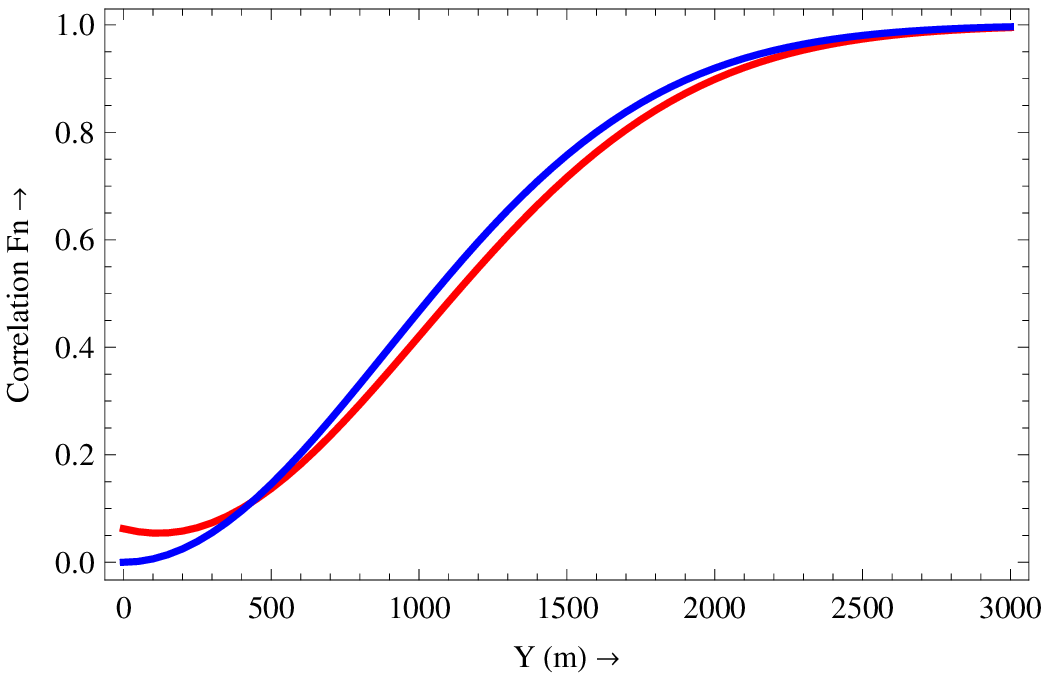}} \\
\subfloat []{ \includegraphics[angle=0,width=2.1in]{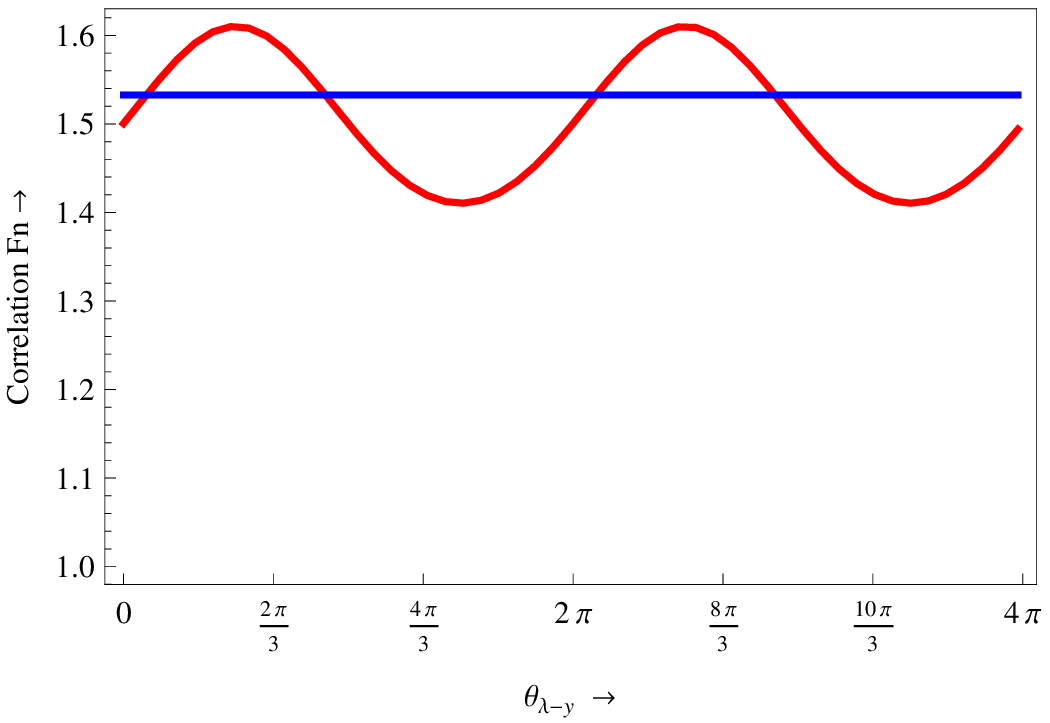}}
 \hspace{.75cm}\vspace{.15cm}
\subfloat []{\includegraphics[angle=0,width=2.1in]{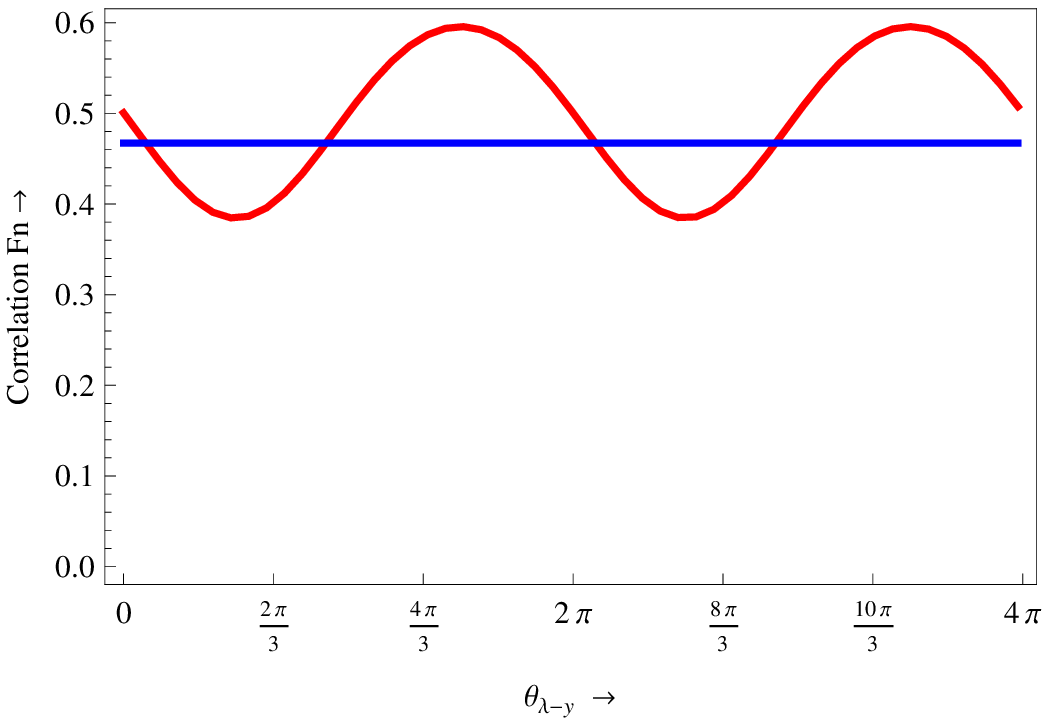}}
\caption{ In the above figures, the blue and red line represent the commutative and noncommutative HBT correlation functions respectively. 
The figures (a), (c) are plotted for ordinary and twisted bosons and (b), (d) are plotted for ordinary and twisted fermions.
The figures (a), (b) are plotted with $|\bold{p_{0}}| = 6 \times 10^{19} $eV (i.e. at G.Z.K cutoff),  $\alpha = 2.04 \times 10^{19}\text{eV}^{-2} $ (i.e taken numerically same
 as the present error in estimation of energy of UHECRs) \cite{nag}, $|\bold{\lambda}| =  1.47 \times 10^{-24}\text{m}^{2}$  and taken along z-axis and the angles as 
 $\theta_{\lambda - y}= \frac{\pi}{4}$, $\theta_{\lambda - p_{0}}= \frac{\pi}{4}$ , $\phi_{\lambda - y}= \frac{\pi}{3}$ and $ \phi_{\lambda - p_{0}}= \frac{\pi}{6} $. 
The figures (c), (d) are plotted with $|\bold{y}|= 1000$m  and same values for rest of the parameters.     }

\end{figure}

Of particular interest are the various angular dependences of noncommutative correlation functions (\ref{gcor}) and (\ref{noncor}) which are completely absent in commutative 
correlation functions (\ref{corgaus}) and (\ref{comcor}). Due to rotation and revolution of earth, the noncommutative correlation functions will show periodic oscillations completely
absent in commutative correlations and hence will perhaps provide the best and most unambiguous signal for underlying noncommutative structure of spacetime. Therefore, as claimed earlier, 
the information about noncommutative structure of spacetime, can be extracted out, from observing the nature of variation of HBT correlation functions (in particular by looking at 
angular variations) with varying certain experimentally measurable quantities, in a very unambiguous way. \\ 

Also it is worth noting that if we take noncommutative length scale same as Planck scale i.e. $10^{-35}$m then due to an upper limit on momenta of UHECRs (GZK cutoff), the deviations 
will turn out to be O($10^{-50}$) which are too small to give any unambiguous signatures of it. So the noncommutative deviations are detectable only if the effective 
noncommutative scale is much larger than Planck scale, which is likely to happen in presence of large extra dimensions. Thus the noncommutative deviations in HBT correlations 
effectively provide us signatures of large extra dimensions.  \\

%%%%%%%%%%%%%%%%%%%%%%%%%%%%%%%%%%%%%%%%%%%%%%%%%%%%%%%%%%%%%%%%%%%%%%%%%%%%%%%%%%%%%%%%%%%%%%%%%%%%%%%%%%%%%%%%%%%%%%%%%%%%%%%%%%%%%%%%%%%%%%%%%%%%%%%%%%%%%%%%%%%%%%%%%%%%%%%%%%%%%%%%

\textbf{Acknowledgements :} I am grateful to Sachindeo Vaidya for suggesting this problem and for many useful discussions and critical comments. 
I am also indebted to Prasad Basu for useful discussions and critical comments on this work. This work was supported by C.S.I.R under the 
award no: F. No 10 2(5)/2007(1) E.U. II.

%%%%%%%%%%%%%%%%%%%%%%%%%%%%%%%%%%%%%%%%%%%%%%%%%%%%%%%%%%%%%%%%%%%%%%%%%%%%%%%%%%%%%%%%%%%%%%%%%%%%%%%%%%%%%%%%%%%%%%%%%%%%%%%%%%%%%%%%%%%%%%%%%%%%%%%%%%%%%%%%%%%%%%%

%%%%%%%%%%%%%%%%%%%%%%%%%%%%%%%%%%%%%%%%%%%%%%%%%%%%%%%%%%%%%%%%%%%%%%%%%%%%%%%%%%%%%%%%%%%%%%%%%%%%%%%%%%%%

\end{document}